\documentclass[aps,prd,superscriptaddress,showpacs]{revtex4}
\usepackage{graphicx}
\usepackage{epstopdf}
\usepackage{amsmath}
\usepackage{amsfonts}
\usepackage{amssymb}
\usepackage{latexsym}

\begin{document}
\title{Remarks on nonlinear electrodynamics II}
\author{Patricio Gaete}\email{patricio.gaete@usm.cl}
 \affiliation{Departmento de F\'{i}sica and Centro Cient\'{i}fico-Tecnol\'ogico de Valpara\'{i}so, Universidad T\'{e}cnica Federico Santa Mar\'{i}a, Valpara\'{i}so, Chile}
\date{\today }

\begin{abstract}
We consider both massive Euler-Heisenberg-like and Euler-Heisenberg-like Electrodynamics in the approximation of the strong-field limit. Our analysis shows that massive Euler-Heisenberg-type Electrodynamics displays the vacuum birefringence phenomenon. 
Afterwards, we calculate the lowest-order modifications to the interaction energy for both classes of Electrodynamics, within the framework of the gauge-invariant path-dependent variables formalism. Interestingly enough, for massive Euler-Heisenbeg-like electrodynamics (Whichmann-Kroll) we obtain a new long-range ($1/r^3$- type) correction,
apart from a long-range ($1/r^5$- type) correction to the Coulomb potential. However, Euler-Heisenberg-like Electrodynamics in the approximation of the strong-field limit (to the leading logarithmic order), display a long-range ($1/r^5$- type) correction to the Coulomb potential. Again, for their non-commutative versions, the interaction energy is ultraviolet finite.

\end{abstract}
\maketitle

\section{Introduction}

The phenomenon of vacuum polarization in Quantum Electrodynamics (QED), arising from the polarization of virtual electron-positron pairs and leading to nonlinear interactions between electromagnetic fields, remain as exciting as in the early days of QED \cite{Breit-Wheeler, Adler, Costantini,Ruffini,Dunne}. An example that illustrates this is the scattering of photons by photons, which despite remarkable progresses has not yet been confirmed  \cite{Bamber,Burke,nphoton,Tommasini1,Tommasini2}. Along the same line, we also recall that alternative scenarios such as Born-Infeld theory \cite{Born}, millicharged particles \cite{Gies} or axion-like particles \cite{Masso,Gaete1,Gaete2} may have more significant contributions to photon-photon scattering physics. 

Interestingly, it should be recalled here that the physical effect of vacuum polarization appears as a modification in the interaction energy between heavy charged particles. In fact, this physical effect changes both the strength and the structural form of the interaction energy. This clearly requires the addition of correction terms in the Maxwell Lagrangian to incorporate the contributions from vacuum polarization process. Two important examples of such a class of contributions are the Uehling and Serber correction and the Wichmann-Kroll correction, which can be derived from the Euler-Heisenberg Lagrangian. Incidentally, as was explained in \cite{Dunne}, it is of interest to notice that the Euler-Heisenberg result extends the Euler-Kockel calculation (in the constant background field limit), which contains nonlinear corrections in powers of the field strengths. Whereas the Uehling and Serber result contains corrections linear in the fields (but nonlinear in the space-time dependence of the background fields). We further mention that, as in the Euler-Heisenbeg case, Born-Infeld (BI) Electrodynamics also contains similar nonlinear corrections to Maxwell theory from a classical point of view, as is well known. Nevertheless, BI Electrodynamics is distinguished, since BI-type effective actions arise in many different contexts in superstring theory \cite{Tseytlin,Pope}. In addition to Born-Infeld theory, other types of nonlinear electrodynamics have been discussed in the literature \cite{Gaete-Helayel1,Gaete-Helayel2,Hendi,Zhao,Olea,Habib}.

In this context, we also point out that extensions of the Standard Model (SM) such as Lorentz invariance violating scenarios and fundamental length have become the focus of intense research activity \cite{AmelinoCamelia:2002wr,Jacobson:2002hd,Konopka:2002tt,Hossenfelder:2006cw,Nicolini:2008aj}. This has its origin in the fact that the SM does not include a quantum theory of gravitation, as to circumvent difficulties theoretical in the quantum gravity program. Within this context quantum field theories allowing non-commuting position operators has been studied by using a start product (Moyal product) \cite{Witten:1985cc,Seiberg:1999vs,Douglas:2001ba,Szabo:2001kg,Gomis:2000sp,Bichl:2001nf}. Mention should be made, at this point, to a novel way to formulate noncommutative quantum field theory (or quantum field theory in the presence of a minimal length)  \cite{Euro1,Euro2,Euro3}, which is implemented through a new multiplication rule which is known as Voros star-product \cite{Banerjee:2009xx}. It was later shown that the physics is independent from the choice of the type of product \cite{Jabbari}. More recently, this new approach has been successfully extended to black holes physics \cite{Piero-Euro}.

Inspired by these observations, the purpose of this paper is to extend our previous studies \cite{Gaete-Helayel1,Gaete-Helayel2} on nonlinear electrodynamics to the case when vacuum polarization corrections are taken into account. The preceding studies were done using the gauge-invariant but path-dependent variables formalism, where the interaction potential energy between two static charges is determined by the geometrical condition of gauge invariance. One important advantage of this approach is that it provides a physically-based alternative to the usual Wilson loop approach. Accordingly, we shall work out the static potential for electrodynamics which include, apart from the Maxwell Lagrangian, additional terms corresponding to the Uehling, massive Euler-Heisenberg-like, Euler-Heisenberg Electrodynamics in the approximation of the strong-field limit (to the leading logarithmic order)
and for their non-commutative versions. Our results show a long-range ${\raise0.7ex\hbox{$1$} \!\mathord{\left/{\vphantom {1 {{L^5}}}}\right.\kern-\nulldelimiterspace}\!\lower0.7ex\hbox{${{L^5}}$}}$-type correction to the Coulomb potential for both massive Euler-Heisenberg-like and Euler-Heisenberg Electrodynamics in the approximation of the strong-field limit (to the leading logarithmic order). Interestingly enough, for massive Euler-Heisenbeg-like electrodynamics (Whichmann-Kroll) we obtain a new long-range ${\raise0.7ex\hbox{$1$} \!\mathord{\left/{\vphantom {1 {{L^3}}}}\right.\kern-\nulldelimiterspace}\!\lower0.7ex\hbox{${{L^3}}$}}$ correction to the interaction energy. Nevertheless, for their non-commutative versions, the static potential becomes ultraviolet finite.

The organization of the paper is as follows: In Section II, we reexamine Uehling Electrodynamics in order to establish a framework for the computation of the static potential. In Section III we consider Euler-Heisenberg-like (with a mass term) and show that it yields birefringence, compute the interaction energy for a fermion-antifermion pair and its version in the presence of a minimal length. In Section III, we repeat our analysis for Euler-Heisenberg Electrodynamics in the approximation of the strong-field limit. Finally, in Section IV, we cast our Final Remarks.

In our conventions the signature of the metric is $(+1,-1,-1,-1)$.

\section{Brief review on the Uehling potential} 

As already expressed, we now reexamine the interaction energy for Maxwell theory with an additional term corresponding to the Uehling correction (Uehling Electrodynamics). This would not only provide the setup theoretical for our subsequent work, but also fix the notation. To do that we will calculate the expectation value of the energy operator $ H$ in the physical state $ |\Phi\rangle$, which we will denote by ${ \langle H\rangle_ \Phi}$. We start off our analysis by considering the effective Lagrangian density \cite{Dittrich}:
\begin{equation}
{\cal L} =  - \frac{1}{4}{F_{\mu \nu }}\left( {1 - \frac{\alpha }{{3\pi }}\Delta {\cal M}} \right){F^{\mu \nu }}, \label{Vac05}
\end{equation}
where 
\begin{equation}
{\cal M}\left( {m,x} \right) = \int_{4{m^2}}^\infty  {d\tau } \frac{1}{{\tau \left( {\tau  + \Delta } \right)}}\left( {1 + \frac{{2{m^2}}}{\tau }} \right)\sqrt {1 - \frac{{4{m^2}}}{\tau }}, \label{Vac06}
\end{equation}
with $\Delta  \equiv {\partial _\mu }{\partial ^\mu }$. It should be noted that $\cal M$ contains the effect of vacuum polarization to first order in the fine structure constant, $\alpha  = \frac{{{e^2}}}{{\hbar c}}$, and, $m$, is the electron mass.

To obtain the corresponding Hamiltonian, we will carry out the quantization of the theory using the Hamiltonian formalism. In this way, the canonical momenta are found to be ${\Pi ^\mu } = \left( {1 - \frac{\alpha }{{3\pi }}\Delta {\cal M}} \right){F^{\mu 0}}$, and one immediately identifies the primary constraint $\Pi ^0  = 0$, whereas the momenta are ${\Pi ^i} = \left( {1 - \frac{\alpha }{{3\pi }}\Delta {\cal M}} \right){E^i}$ (with ${E_i} = {F_{i0}}$). Accordingly, the canonical Hamiltonian $H_C$ is
\begin{equation}
{H_C} = \int {{d^3}x} \left[ {{\Pi ^i}{\partial _i}{A_0} - \frac{1}{2}{\Pi _i}{{\left( {1 - {\textstyle{\alpha  \over {3\pi }}}\Delta {\cal M}} \right)}^{ - 1}}{\Pi ^i} + \frac{1}{4}{F_{ij}}\left( {1 - {\textstyle{\alpha  \over {3\pi }}}\Delta {\cal M}} \right){F^{ij}}} \right]. \label{Vac10}
\end{equation}
As usual, requiring the primary constraint $\Pi^{0}$ to be stationary, leads to the secondary constraint $\Gamma _1  \equiv \partial _i \Pi ^i  = 0$. It is straightforward to check that there are no further constraints in the theory. Hence, the extended Hamiltonian that generates translations in time then reads $  
H = H_C  + \int {d^3 x} \left( {u_0(x) \Pi_0(x)  + u_1(x) \Gamma _1(x) } \right)$, where $u_o(x)$ and $u_1(x)$ are arbitrary Lagrange multipliers. With the aid of equation (\ref{Vac10}) we find that $\dot A_0 \left( x \right) = \left[ {A_0 \left( x \right),H} \right] = u_0 \left( x \right)$, which is an arbitrary function. Since $\Pi^0=0$ always, neither $A^0$ nor $\Pi^0$ are of interest in describing the system and may be discarded from the theory. Therefore, the Hamiltonian becomes
\begin{equation}
{H} = \int {{d^3}x} \left[ {w(x){\partial _i}\Pi ^i - \frac{1}{2}{\Pi _i}{{\left( {1 - {\textstyle{\alpha  \over {3\pi }}}\Delta {\cal M}} \right)}^{ - 1}}{\Pi ^i} + \frac{1}{4}{F_{ij}}\left( {1 - {\textstyle{\alpha  \over {3\pi }}}\Delta {\cal M}} \right){F^{ij}}} \right], \label{Vac15}
\end{equation}
where $w(x) = u_1 (x) - A_0 (x)$.

According to the usual procedure, we impose a gauge condition such that the full set of constraints becomes second class. A convenient choice is
\begin{equation}
\Gamma _2 \left( x \right) \equiv \int\limits_{C_{\xi x} } {dz^\nu
} A_\nu \left( z \right) \equiv \int\limits_0^1 {d\lambda x^i }
A_i \left( {\lambda x} \right) = 0. \label{Vac20}
\end{equation}
where  $\lambda$ $(0\leq \lambda\leq1)$ is the parameter describing
the spacelike straight path $ z^i = \xi ^i  + \lambda \left( {x -
\xi } \right)^i $, and $ \xi $ is a fixed point (reference point).
There is no essential loss of generality if we restrict our
considerations to $ \xi ^i=0 $. The Dirac brackets can now be determined and we simply note the only nontrivial Dirac bracket involving the canonical variables, that is,
\begin{equation}
\left\{ {A_i \left( x \right),\Pi ^j \left( y \right)} \right\}^ *
= \delta _i^j \delta ^{\left( 3 \right)} \left( {x - y} \right) -
\partial _i^x \int\limits_0^1 {d\lambda x^i } \delta ^{\left( 3
\right)} \left( {\lambda x - y} \right). \label{Vac25}
\end{equation}
In passing we also recall that the transition to a quantum theory is made by the replacement of the Dirac brackets by the operator commutation relations according to ${\left\{ {A,B} \right\}^ * } \to \left( { - {\raise0.5ex\hbox{$\scriptstyle i$}
\kern-0.1em/\kern-0.15em
\lower0.25ex\hbox{$\scriptstyle \hbar $}}} \right)\left[ {A,B} \right]$. 

With the foregoing information, we can now proceed to obtain the interaction energy. As already mentioned, in order to accomplish this purpose, we will calculate the expectation value of the energy operator $H$ in the physical state $\left| \Phi  \right\rangle$, where the physical states $\left| \Phi  \right\rangle$ are gauge-invariant ones. The physical state can be written as 
\begin{equation}
\left| \Phi  \right\rangle  \equiv \left| {\bar \Psi ({\bf y})\Psi ({{\bf y}^ \prime })} \right\rangle  = \bar \psi ({\bf y})\exp (\frac{{iq}}{\hbar }\int_{{{\bf y}^ \prime }}^{\bf y} {d{z^i}{A_i}(z)} )\psi ({{\bf y}^ \prime })\left| 0 \right\rangle,                              
\label{Vac30}
\end{equation}
where $\left| 0 \right\rangle$ is the physical vacuum state and the line integral appearing in the above expression is along a space-like path starting at ${\bf y}\prime$ and ending at $\bf y$, on a fixed
time slice. 

Making use of the above Hamiltonian structure \cite{Gaete-Helayel1}, we find that 
\begin{equation}
\Pi _i \left( x \right)\left| {\overline \Psi \left( \mathbf{y }\right)\Psi
\left( {\mathbf{y}^ \prime } \right)} \right\rangle = \overline \Psi \left( 
\mathbf{y }\right)\Psi \left( {\mathbf{y}^ \prime } \right)\Pi _i \left( x
\right)\left| 0 \right\rangle + q\int_ {\mathbf{y}}^{\mathbf{y}^ \prime } {\
dz_i \delta ^{\left( 3 \right)} \left( \mathbf{z - x} \right)} \left| \Phi
\right\rangle.  \label{Vac35}
\end{equation}
With the aid of equations (\ref{Vac35}) and (\ref{Vac15}), the lowest-order modification in ${\alpha}$ of the interaction energy takes the form
\begin{equation}
{\left\langle H \right\rangle _\Phi } = {\left\langle H \right\rangle _0} + {V_1} + {V_2}, \label{Vac40}
\end{equation}
where $\left\langle H \right\rangle _0  = \left\langle 0 \right|H\left| 0 \right\rangle$. The $V_1$, $V_2$ terms are given by
\begin{equation}
{V_1} = \frac{{{q^2}}}{2}\int {{d^3}x} \int_{{{\bf y}^ \prime }}^{\bf y} {d{z^i}} {\delta ^{\left( 3 \right)}}\left( {{\bf x} - {\bf z}} \right)\int_{{{\bf y}^ \prime }}^{\bf y} {d{z^ \prime}^i} {\delta ^{\left( 3 \right)}}\left( {{\bf x} - {{\bf z}^ \prime }} \right), \label{Vac45} 
\end{equation}
and
\begin{equation}
{V_2} = \frac{{{q^2}}}{2}\frac{\alpha }{{3\pi }}\int_{4{m^2}}^\infty  {d\tau } \frac{{\rho \left( \tau  \right)}}{\tau }\int {{d^3}} x\int_{\bf y}^{{{\bf y}^ \prime }} {dz_i^ \prime } {\delta ^{\left( 3 \right)}}\left( {{{\bf z}^ \prime } - {\bf x}} \right)\nabla _x^2\frac{1}{{\tau  - \nabla _x^2}}\int_{\bf y}^{{{\bf y}^ \prime }} {d{z^i}} {\delta ^{\left( 3 \right)}}\left( {{\bf z} - {\bf x}} \right), \label{Vac50}
\end{equation}
where $\rho \left( \tau  \right) = \left( {1 + \frac{{2{m^2}}}{\tau }} \right)\sqrt {1 - \frac{{4{m^2}}}{\tau }}$.

We note that the term (\ref{Vac45}) may look peculiar, but it is nothing but the familiar Coulomb interaction plus a self-energy term \cite{PG2}. Now making use of the Green function, $G\left( {{\bf z},{{\bf z}^ \prime }} \right) = \frac{1}{{4\pi }}\frac{{{e^{ - \sqrt \tau  |{\bf z} - {{\bf z}^ \prime }|}}}}{{|{\bf z} - {{\bf z}^ \prime }|}}$, the term (\ref{Vac50}) can be rewritten in the form
\begin{equation}
{V_2} = \frac{{{q^2}}}{2}\frac{\alpha }{{3\pi }}\int_{4{m^2}}^\infty  {d\tau } \frac{{\rho \left( \tau  \right)}}{\tau }\int_{\bf y}^{{{\bf y}^ \prime }} {dz_i^ \prime } \nabla _{{z^ \prime }}^2\int_{\bf y}^{{{\bf y}^ \prime }} {d{z^i}} G\left( {{\bf z},{{\bf z}^ \prime }} \right) = - \frac{\alpha }{{3\pi }}\frac{{{q^2}}}{2}\int_{4{m^2}}^\infty  {d\tau } \frac{{\rho \left( \tau  \right)}}{\tau }\frac{{{e^{ - \sqrt \tau  |{\bf y} - {{\bf y}^ \prime }|}}}}{{|{\bf y} - {{\bf y}^ \prime }|}}
. \label{Vac55}
\end{equation}

Since the second and third term on the right-hand side of Eq. (\ref{Vac40}) are clearly dependent on the distance between the external static fields, the potential for two opposite charges located at ${\bf y}$ and ${{\bf y}\prime}$ reads
\begin{equation}
V =  - \frac{{{q^2}}}{{4\pi }}\frac{1}{L}\left( {1 + \frac{\alpha }{{3\pi }}\int_{4{m^2}}^\infty  {d\tau } \frac{{\rho \left( \tau  \right)}}{\tau }{e^{ - \sqrt \tau  L}}} \right),
\label{Vac60}
\end{equation}
where $L = |{\bf y} - {{\bf y}^ \prime }|$. Accordingly, one recovers the known Uehling potential, which finds here an entirely different derivation.

Before we proceed further, we wish to show that this result can be written alternatively in a more explicit form. Making use of \cite{Klarsfield}
\begin{equation}
{\chi _n}\left( z \right) = \int_1^\infty  {dt} \frac{{{e^{ - tz}}}}{{{t^n}}}\left( {1 + \frac{1}{{2{t^2}}}} \right)\sqrt {1 - \frac{1}{{{t^2}}}}, \label{Vac65}
\end{equation}
we then get
\begin{equation}
V =  - \frac{{{q^2}}}{{4\pi }}\frac{1}{L}\left( {1 + \frac{{2\alpha }}{{3\pi }}{\chi _1}\left( {2mL} \right)} \right). \label{Vac75}
\end{equation}
By the transformation, $t = \cosh u$ \cite{Frolov}, the functions ${\chi _n}$ can be reduced to the form \cite{Klarsfield}:
\begin{equation}
{\chi _n}\left( z \right) = K{i_{n - 1}}\left( z \right) - \frac{1}{2}K{i_{n + 1}}\left( z \right) - \frac{1}{2}K{i_{n + 3}}\left( z \right), \label{Vac70}
\end{equation}
where the functions $Ki$ denote Bessel function integrals. Hence we see that the interaction energy (with $m=1$) becomes
\begin{equation}
V =  - \frac{{{q^2}}}{{4\pi }}\frac{1}{L}\left\{ {1 + \frac{{2\alpha }}{{3\pi }}\left[ {\left( {1 + \frac{{{L^2}}}{3}} \right){K_0}\left( {2L} \right) - \left( {\frac{{5L}}{3} + \frac{{2{L^3}}}{3}} \right){K_1}\left( {2L} \right) + \left( {\frac{{3L}}{2} + \frac{{2{L^3}}}{3}} \right)\int_{2L}^\infty  {dt{K_0}\left( t \right)} } \right]} \right\}, \label{Vac75}
\end{equation}
where $K_0(z)$ and $K_1(z)$ are modified Bessel functions. Finally, with the aid of asymptotic forms for Bessel functions, it is a simple matter to find expressions for $V$ for large and small $L$.

Before concluding this subsection we discuss an alternative way of stating our previous result (\ref{Vac60}), which displays certain distinctive features of our methodology. We start by considering \cite{PG2,PG1}
\begin{equation}
V \equiv q\left( {{\cal A}_0 \left( {\bf 0} \right) - {\cal A}_0 \left( {\bf L} \right)} \right), \label{Vac80}
\end{equation}
where the physical scalar potential is given by
\begin{equation}
{\cal A}_0 (t,{\bf r}) = \int_0^1 {d\lambda } r^i E_i (t,\lambda
{\bf r}). \label{Vac85}
\end{equation}
This follows from the vector gauge-invariant field expression
\begin{equation}
{\cal A}_\mu  (x) \equiv A_\mu  \left( x \right) + \partial _\mu  \left( { - \int_\xi ^x {dz^\mu  A_\mu  \left( z \right)} } \right), \label{Vac90}
\end{equation}
where the line integral is along a space-like path from the point $\xi$ to $x$, on a fixed slice time. It is also important to observe that the gauge-invariant variables (\ref{Vac85}) commute with the sole first constraint (Gauss law), showing in this way that these fields are physical variables. Inasmuch as we are interested in estimating the lowest-order correction to the Coulomb energy, we will retain only the leading term in expression
${E^i} = {\left( {1 - \frac{\alpha }{{3\pi }}\Delta {\cal M}} \right)^{ - 1}}{\Pi ^i}$. Making use of this last expression, equation (\ref{Vac85}) gives
\begin{equation}
{{\cal A}_0}(t,{\bf r}) = \int_0^1 {d\lambda } {r^i}{\partial _i}{\left( { - \frac{{{J^0}}}{{{\nabla ^2}}}} \right)_{\lambda r}} + \frac{\alpha }{{3\pi }}\int_{4{m^2}}^\infty  {\frac{{d\tau }}{\tau }} \rho \left( \tau  \right)\int_0^1 {d\lambda } {r^i}{\partial _i}{\left( { - \frac{{{J^0}}}{{{\nabla ^2} - \tau }}} \right)_{\lambda r}}, \label{Vac95}
\end{equation}
to get the last line we used Gauss law for the present theory, that is, $\partial _i \Pi ^i  = J^0$ (where we have included the external current $J^0$ to represent the presence of two opposite charges). Accordingly, for $J^0 (t,{\bf r}) = q\delta ^{\left( 3 \right)} \left( {\bf r} \right)$, the potential for a pair of static point-like opposite charges located at $\bf 0$ and $\bf L$, is given by
\begin{equation}
V =  - \frac{{{q^2}}}{{4\pi }}\frac{1}{L}\left( {1 + \frac{\alpha }{{3\pi }}\int_{4{m^2}}^\infty  {d\tau } \frac{{\rho \left( \tau  \right)}}{\tau }{e^{ - \sqrt \tau  L}}} \right),
\label{Vac100}
\end{equation}
after substracting a self-energy term.

\section{Euler-Heisenberg-like model}

Proceeding in the same way as we did in the foregoing section, we shall now consider the interaction energy for Euler-Heisenberg-like electrodynamics. Nevertheless, in order to put our discussion into context it is useful to describe very briefly the model under consideration. In such a case the Lagrangian density reads:
\begin{equation}
{\cal L} = \frac{{{\beta ^2}}}{2}\left\{ {1 - {{\left[ {1 + \frac{1}{{{\beta ^2}}}{\cal F} - \frac{1}{{{\beta ^2}{\gamma ^2}}}{{\cal G}^2}} \right]}^p}} \right\},  \label{E-H05}
\end{equation}
where have included two parameters $\beta$ and $\gamma$. As usual, ${\cal F} = \frac{1}{4}F_{\mu \nu } F^{\mu \nu }$,    
${\cal G} = \frac{1}{4}F_{\mu \nu } \tilde F^{\mu \nu }$, $F_{\mu \nu }  = \partial _\mu  A_\nu   - \partial _\nu  A_\mu$
and $\tilde F^{\mu \nu }  = \frac{1}{2}\varepsilon ^{\mu \nu \rho \lambda } F_{\rho \lambda }$. Let us also mention here that in our previous paper \cite{Gaete-Helayel2} we have studied the domain $0 < p < 1$. Moreover, follows from (\ref{E-H05}) that when $p=2$ the model contains, to order ${\cal O}\left( {{\raise0.5ex\hbox{$\scriptstyle 1$}
\kern-0.1em/\kern-0.15em\lower0.25ex\hbox{$\scriptstyle {{\beta ^2}}$}}} \right)$ and ${\cal O}\left( {{\raise0.5ex\hbox{$\scriptstyle 1$}\kern-0.1em/\kern-0.15em\lower0.25ex\hbox{$\scriptstyle {{\gamma ^2}}$}}} \right)$, a Euler-Heisenberg-like model with the appropriate identifications of the constants. Interestingly, we also observe that in the limit $\gamma  \to \infty$ we obtain a Whichmann-Kroll model. This remark opens up the way to discuss the effect of these nonlinear corrections on the interaction energy, as we are going to study below. In fact, we shall consider a massive Whichmann-Kroll system. The motivation for this study comes from recent considerations in the context of dualities \cite{Ferrara}, where massive Born-Infeld systems play an important role.  

Having made these observations we can write immediately the field equations for $p=2$:
\begin{equation}
\partial _\mu  \left[ {{\Gamma }\left( {F^{\mu \nu }  - \frac{2}{{\gamma ^2 }}{\cal G}\tilde F^{\mu \nu } } \right)} \right] = 0, \label{E-H10}
\end{equation}
while the Bianchi identities are given by
\begin{equation}
\partial _\mu  \tilde F^{\mu \nu }  = 0, \label{E-H15}
\end{equation}
where
\begin{equation}
\Gamma  = 1 + \frac{{{\cal F}}}{{\beta ^2 }} - \frac{{{\cal G}^2 }}{{\beta ^2 \gamma ^2}}. \label{E-H20}
\end{equation}
Also, it is straightforward to see that Gauss law becomes,
\begin{equation}
\nabla  \cdot {\bf D} = 0, \label{E-H25}
\end{equation}
where $\bf D$ is given by
\begin{equation}
{\bf D} = \left[ {1 - \frac{{\left( {{{\bf E}^2} - {{\bf B}^2}} \right)}}{{{2\beta ^2}}} - \frac{{{{\left( {{\bf E} \cdot {\bf B}} \right)}^2}}}{{{\beta ^2 \gamma ^2}}}} \right]\left( {{\bf E} + \frac{2}{{{\gamma ^2}}}\left( {{\bf E} \cdot {\bf B}} \right){\bf B}} \right). \label{E_H30}
\end{equation}
Again, from equation (\ref{E-H25}), for $J^0 (t,{\bf r}) = e\delta ^{\left( 3 \right)} \left( {\bf r} \right)$, we find ${\bf D} = \frac{Q}{{r^2 }}\hat r$, where $Q = \frac{e}{{4\pi }}$. This then implies that for a point-like charge, e, at the origin, the expression
\begin{equation}
\frac{Q}{{{r^2}}} = \left( {1 - \frac{{{{\bf E}^2}}}{{{2\beta ^2}}}} \right)|{\bf E}|, \label{EH-35}
\end{equation}
tells us that, for $r \to 0$, the electrostatic field becomes singular at $r=0$, in contrast to the $0 < p < 1$ case where the electrostatic field is finite. Even so, in this theory the phenomenon of birefringence is present.  

To illustrate this important feature we introduce the vectors ${\bf D} = {{\partial {\cal L}} \mathord{\left/
 {\vphantom {{\partial L} {\partial {\bf E}}}} \right.
 \kern-\nulldelimiterspace} {\partial {\bf E}}}$ and ${\bf H} =  - {{\partial {\cal L}} \mathord{\left/
 {\vphantom {{\partial L} {\partial {\bf B}}}} \right.
 \kern-\nulldelimiterspace} {\partial {\bf B}}}$:
\begin{equation}
{\bf D} = \Gamma \left( {{\bf E} + 2\frac{{{\bf B}\left( {{\bf E} \cdot {\bf B}} \right)}}{{{\gamma ^2}}}} \right),  \label{E-H40}
\end{equation}
and
\begin{equation}
{\bf H} = \Gamma \left( {{\bf B} - 2\frac{{{\bf E}\left( {{\bf E} \cdot {\bf B}} \right)}}{{{\gamma ^2}}}} \right),  \label{E-H45}
\end{equation}
where $ 
\Gamma  = 1 + \frac{1}{{2\beta ^2 }}\left( {{\bf B}^2  - {\bf E}^2 } \right) - \frac{1}{{\beta ^2 \gamma ^2 }}\left( {{\bf E} \cdot {\bf B}} \right)^2$. We thus obtain the equations of motion 
\begin{equation}
\nabla  \cdot {\bf D} = 0, \  \  \
\frac{{\partial {\bf D}}}{{\partial t}} - \nabla  \times {\bf H} = 0, \label{E-H50}
\end{equation}
and
\begin{equation}
\nabla  \cdot {\bf B} = 0, \  \  \
\frac{{\partial {\bf B}}}{{\partial t}} + \nabla  \times {\bf E} = 0. \label{E-H55}
\end{equation}
With the aid from  equations (\ref{E-H40}) and (\ref{E-H45}) we find the electric permitivity, 
$\varepsilon _{ij}$, and the inverse magnetic permeability, $\left( {\mu}^{-1}  \right)_{ij}$, tensors of the vacuum, that is,
\begin{equation}
{\varepsilon _{ij}} = \Gamma \left( {{\delta _{ij}} + \frac{{2{B_i}{B_j}}}{{{\gamma ^2}}}} \right), \ \ \
{\left( {{\mu ^{ - 1}}} \right)_{ij}} = \Gamma \left( {{\delta _{ij}} - \frac{{2{E_i}{E_j}}}{{{\gamma ^2}}}} \right)\ , \label{E-H60}
\end{equation}
with $D_i  = \varepsilon _{ij} E_j$ and $B_i  = \mu _{ij} H_j $.

In accordance with our previous procedure \cite{Gaete-Helayel1,Gaete-Helayel2}, we can now linearize the above equations. To do this, it is advantageous to introduce a weak electromagnetic wave $({\bf E_p}, {\bf B_p})$ propagating in the presence of a strong constant external field $({\bf E_0}, {\bf B_0})$. On these assumptions, we readily find that, for the case of a purely magnetic field (${\bf E_0}=0$), the vectors ${\bf D}$ and ${\bf H}$ become   
\begin{equation}
{\bf D} = \left( {1 + \frac{{{{\bf B}_0^2}}}{{{2\beta ^2}}}} \right)\left[ {{{\bf E}_p} + \frac{2}{{{\gamma ^2}}}\left( {{{\bf E}_p} \cdot {{\bf B}_0}} \right){{\bf B}_0}} \right],
\end{equation} \label{E-H65}
and
\begin{equation}
{\bf H} = \left( {1 + \frac{{{\bf B}_0^2}}{{{2\beta ^2}}}} \right)\left[ {{{\bf B}_p} + \frac{1}{{{\beta ^2}\left( {1 + \frac{{{\bf B}_0^2}}{{{2\beta ^2}}}} \right)}}\left( {{{\bf B}_p} \cdot {{\bf B}_0}} \right){{\bf B}_0}} \right], \label{E-H70} 
\end{equation}
where we have keep only linear terms in ${\bf E_p}$, ${\bf B_p}$. As before, we consider the $z$ axis as the direction of the external magnetic field (${\bf B_0}  = B_0 {\bf e}_3$) and assuming that the light wave moves along the $x$ axis, the decomposition into a plane wave for the fields ${\bf E}_p$ and ${\bf B}_p$ can be written as
\begin{equation}
{{\bf E_p}}\left( {{\bf x}
,t} \right) = {\bf E}
{e^{ - i\left( {wt - {\bf k} \cdot {\bf x}} \right)}}, \ \ \
{{\bf B_p}}\left( {{\bf x},t} \right) = {\bf B}{e^{ - i\left( {wt - {\bf k} \cdot {\bf x}} \right)}}. \label{E-H75}
\end{equation}
In this case, it clearly follows that 
\begin{equation}
\left( {\frac{{{k^2}}}{{{w^2}}} - {\varepsilon _{22}}{\mu _{33}}} \right){E_2} = 0, \label{E-H80}
\end{equation}  
and
\begin{equation}
\left( {\frac{{{k^2}}}{{{w^2}}} - {\varepsilon _{33}}{\mu _{22}}} \right){E_3} = 0. \label{E-H85}
\end{equation}

As a consequence, we have two different situations: First, if ${\bf E}\ \bot \ {\bf B}_0$ (perpendicular polarization), from (\ref{E-H85}) $E_3=0$, and from (\ref{E-H80}) we get $\frac{{{k^2}}}{{{w^2}}} = {\varepsilon _{22}}{\mu _{33}}$. This then means that the dispersion relation of the photon takes the form
\begin{equation}
{n_ \bot } = \sqrt {\frac{{1 + \frac{{{\bf B}_0^2}}{{2{\beta ^2}}}}}{{1 + \frac{{3{\bf B}_0^2}}{{2{\beta ^2}}}}}}.  \label{E-H90}
\end{equation}
Second, if ${\bf E}\ || \ {\bf B}_0$ (parallel polarization), from (\ref{E-H80}) $E_2=0$, and from (\ref{E-H85}) we get $\frac{{{k^2}}}{{{w^2}}} = {\varepsilon _{33}}{\mu _{22}}$. This leads to
\begin{equation}
n_\parallel   = \sqrt {1 + \frac{{{2\bf B}_0^2 }}{{\gamma ^2 }}}.  \label{E-H95}
\end{equation}
Thus we verify that in the case of a generalized Euler-Heisenberg electrodynamics the phenomenon of birefringence is present.

We now pass to the calculation of the interaction energy between static point-like sources for a massive Whichman-Kroll-like model, our analysis follows closely that of references \cite{Gaete-Helayel1,Gaete-Helayel2}. The corresponding theory is governed by the Lagrangian density 
\begin{equation}
{\cal L} = - \frac{1}{4}F_{\mu \nu }^2 + \frac{1}{{32{\beta ^2}}}{\left( {{F_{\mu \nu }}{F^{\mu \nu }}} \right)^2} + \frac{{{m^2}}}{2}{A_\mu }{A^\mu }. 
\label{E-H100}
\end{equation}
Next, in order to handle the second term on the right hand in (\ref{E-H100}), we introduce an auxiliary field $\xi$ such that its equation of motion gives back the original theory. This allows us to write the Lagrangian density as
\begin{equation}
{\cal L} =  - \frac{1}{4}{F_{\mu \nu }}{F^{\mu \nu }} + \frac{\xi }{{32{\beta ^2}}}{F_{\mu \nu }}{F^{\mu \nu }} - \frac{1}{{128{\beta ^2}}}{\xi ^2} + \frac{{{m^2}}}{2}{A_\mu }{A^\mu }. \label{E-H105}
\end{equation}
With the redefinition $\eta  = 1 - \frac{\xi }{{8{\beta ^2}}}$, equation (\ref{E-H105}) becomes
\begin{equation}
{\cal L} =  - \frac{1}{4}\eta {F_{\mu \nu }}{F^{\mu \nu }} - \frac{1}{2}{\left( {1 - \eta } \right)^2} + \frac{{{m^2}}}{2}{A_\mu }{A^\mu }. \label{E-H110}
\end{equation}

Before we proceed to work out explicitly the interaction energy, we shall first restore the gauge invariance in equation (\ref{E-H110}). Following an earlier procedure, we readily verify that the canonical momenta read ${\Pi ^\mu } =  - \eta {F^{0\mu }}$, which results in the usual primary constraint $\Pi^0=0$, and ${\Pi ^i} = \eta {F^{i0}}$. In this way one obtains 
\begin{equation}
H_{C} = \int {{d^3}x} \left\{ {{\Pi ^i}{\partial _i}{A_0} + \frac{1}{{2\eta }}{{\bf \Pi} ^2} + \frac{\eta }{2}{{\bf B}^2} - \frac{{{m^2}}}{2}{A_\mu }{A^\mu } + \frac{1}{2}{{\left( {1 - \eta } \right)}^2}} \right\}.  \label{E-H115}
\end{equation}
The consistency condition, $\dot \Pi  = 0$, leads to the constraint $\Gamma  \equiv {\partial _i}{\Pi ^i} + {m^2}A^0 = 0$. As a result, both constraints are second class. To convert the second class system into first class we shall adopt the procedure described previously. Thus, we enlarge the original phase space by introducing a canonical pair of fields $ \theta$ and $\Pi_{\theta}$. It follows, therefore, that a new set of first class constraints can be defined in this extended space: ${\Lambda _1} \equiv {\Pi _0} + {m^2}\theta  = 0$  and ${\Lambda _2} \equiv \Gamma  + {\Pi _\theta } = 0$. This then shows that the new constraints are first class and, therefore, restore the gauge symmetry. From this, the new effective Lagrangian density, after integrating out the $\theta$ fields, becomes
\begin{equation}
{\cal L} =  - \frac{1}{4}{F_{\mu \nu }}\left( {\eta  + \frac{{{m^2}}}{\Delta }} \right){F^{\mu \nu }} - \frac{1}{2}{\left( {1 - \eta } \right)^2}. \label{E-H120}
\end{equation}
 Now, writting $\sigma  = \eta  + \frac{{{m^2}}}{\Delta }$, the expression (\ref{E-H120}) can be brought to the form
\begin{equation}
{\cal L} =  - \frac{1}{4}{F_{\mu \nu }}\sigma {F^{\mu \nu }} - \frac{k}{{128}}{\left( {1 - \sigma  + \frac{{{m^2}}}{\Delta }} \right)^2}, \label{E-H125}
\end{equation}
 where $k = 64{\beta ^2}$.
 
We are now ready to compute the interaction energy. In this case, the canonical momenta are ${\Pi ^\mu } =  - \sigma {F^{0\mu }}$, with the usual primary constraint $\Pi^0=0$, and ${\Pi ^i} = \sigma{F^{i0}}$. Hence the canonical Hamiltonian is expressed as
\begin{equation}
{H_C} = \int {{d^3}x} \left\{ {{\Pi ^i}{\partial _i}{A_0} + \frac{1}{{2\sigma }}{{\bf \Pi} ^2} + \frac{\sigma }{2}{{\bf B}^2} + \frac{k}{{128}}{{\left( {1 - \sigma  + \frac{{{m^2}}}{\Delta }} \right)}^2}} \right\}. \label{E-H130}
\end{equation}
Time conservation of the primary constraint $\Pi^{0}$ yields the secondary constraint $\Gamma _1 \equiv \partial _i \Pi ^i  = 0$. Similarly for the ${\cal P}_\sigma$ constraint yields no further constraints and just determines the field $\sigma$. In this case, at leading order in $\beta$, the field $\sigma$ is given by 
\begin{equation}
\sigma  = \left( {1 + \frac{{{m^2}}}{\Delta } - \frac{{{{\bf B}^2}}}{{2{\beta ^2}}}} \right)\left[ {1 - \frac{3}{{2{\beta ^2}}}\frac{1}{{{{\left( {1 + \frac{{{m^2}}}{\Delta } - \frac{{{{\bf B}^2}}}{{2{\beta ^2}}}} \right)}^3}}}{{\bf \Pi} ^2}} \right], \label{E-H135}
\end{equation}
which will be used to eliminate $\sigma$. As before, the corresponding total (first-class) Hamiltonian that generates the time evolution of the dynamical variables is $  
H = H_C  + \int {d^3 x} \left( {u_0(x) \Pi_0(x)  + u_1(x) \Gamma _1(x) } \right)$, where $u_o(x)$ and $u_1(x)$ are the Lagrange multiplier utilized to implement the constraints. 

In the same way as was done in the previous subsection, the expectation value of the energy operator $H$ in the physical state $\left| \Phi  \right\rangle$ becomes
  \begin{equation}
{\left\langle H \right\rangle _\Phi } = \left\langle \Phi  \right|\int {{d^3}x\left\{ {\frac{1}{2}{\Pi ^i}{{\left( {1 + \frac{{{m^2}}}{\Delta }} \right)}^{ - 1}}{\Pi ^i} + \frac{{15}}{{8{\beta ^2}}}{{\bf \Pi} ^4} - \frac{{15{m^2}}}{{2{\beta ^2}}}{{\bf \Pi} ^2}\frac{1}{\Delta }{{\bf \Pi} ^2}} \right\}} \left| \Phi  \right\rangle, \label{E-H145}
 \end{equation}
 in this last line we have considered only quadratic terms in $m^2$.
 
In such a case, by employing (\ref{E-H145}), the lowest-order modification in ${\beta ^2}$ and $m^2$ of the interaction energy takes the form
\begin{equation}
{\left\langle H \right\rangle _\Phi } = {\left\langle H \right\rangle _0} + {V_1} + {V_2} + {V_3}, \label{E-H150}
\end{equation}
where $\left\langle H \right\rangle _0  = \left\langle 0 \right|H\left| 0 \right\rangle$. The $V_1$, $V_2$ and $V_3$ terms are given by
\begin{equation}
{V_1} = \frac{{{q^2}}}{2}\int {{d^3}x} \int_{{{\bf y}^ \prime }}^{\bf y} {d{z^{ \prime i}}} {\delta ^{\left( 3 \right)}}\left( {{\bf x} - {{\bf z}^ \prime }} \right){\left( {1 - \frac{{{m^2}}}{{{\nabla ^2}}}} \right)^{ - 1}}\int_{{{\bf y}^ \prime }}^{\bf y} {d{z^i}} {\delta ^{\left( 3 \right)}}\left( {{\bf x} - {\bf z}} \right),  \label{E-H150}
\end{equation}
\begin{equation}
{V_2} =  - \frac{{{15q^4}}}{{{8\beta ^2}}}\int {{d^3}x} \int_{{{\bf y}^ \prime }}^{\bf y} {d{z_i}} {\delta ^{\left( 3 \right)}}\left( {{\bf x} - {\bf z}} \right)\int_{{{\bf y}^ \prime }}^{\bf y} {d{z^ \prime }^i} {\delta ^{\left( 3 \right)}}\left( {{\bf x} - {{\bf z}^ \prime }} \right)\int_{{{\bf y}^ \prime }}^{\bf y} {d{u ^k}} {\delta ^{\left( 3 \right)}}\left( {{\bf x} - {\bf u} } \right)\int_{{{\bf y}^ \prime }}^{\bf y} {d{v _k}} {\delta ^{\left( 3 \right)}}\left( {{\bf x} - {\bf v} } \right), \label{E-H155} 
\end{equation}
and
\begin{equation}
{V_3} = \frac{{15{m^2}{q^4}}}{{2{\beta ^2}}}\int {{d^3}x} \int_{{{\bf y}^ \prime }}^{\bf y} {d{z_i}} {\delta ^{\left( 3 \right)}}\left( {{\bf x} - {\bf z}} \right)\int_{{{\bf y}^ \prime }}^{\bf y} {d{z^{ \prime i}}} {\delta ^{\left( 3 \right)}}\left( {{\bf x} - {{\bf z}^ \prime }} \right)\int_{{{\bf y}^ \prime }}^{\bf y} {d{ u ^k}} {\delta ^{\left( 3 \right)}}\left( {{\bf x} - {\bf u} } \right)\int_{{{\bf y}^ \prime }}^{\bf y} {d{v _k}} {\delta ^{\left( 3 \right)}}\left( {{\bf x} - {\bf v}} \right). \label{E-H160} 
\end{equation}

Finally, with the aid of the expressions (\ref{E-H150}), (\ref{E-H155}) and (\ref{E-H160}), the potential for a pair of static point-like opposite charges located at $\bf 0$ and $\bf L$, is given by
\begin{equation}
V =  - \frac{{{q^2}}}{{4\pi }}\frac{{{e^{ - mL}}}}{L} + \frac{{{q^4}}}{{16\pi {\beta ^2}}}\left( {\frac{3}{{8\pi }}\frac{1}{{{L^2}}} - 5{m^2}} \right)\frac{1}{{{L^3}}}. \label{E-H165}
\end{equation}
observe that when $m=0$, the profile (\ref{E-H165}) reduces to the known Whichmann-Kroll interaction energy. On the other hand, for $m \ne 0$, it should be noted the key role played by the mass term in transforming the Coulomb potential into the Yukawa one. Interestingly enough, an unexpected feature is found. In fact, the profile (\ref{E-H165}) displays a new long-range ${\raise0.7ex\hbox{$1$} \!\mathord{\left/{\vphantom {1 {{L^3}}}}\right.\kern-\nulldelimiterspace}\!\lower0.7ex\hbox{${{L^3}}$}}$ correction, where its strength is proportional to ${m^2}$. It is also important to observe that a analogous correction has been found in Born-Infeld electrodynamics in  the context of very special relativity \cite{Bufalo}. In this way we establish a new connection between nonlinear effectives theories.

Before we proceed further, we should comment on our result. In the case of QED (Euler-Heisenberg Lagrangian density) the parameter ${\raise0.7ex\hbox{$1$} \!\mathord{\left/{\vphantom {1 {{\beta ^2}}}}\right.\kern-\nulldelimiterspace}\!\lower0.7ex\hbox{${{\beta ^2}}$}}$ is given by $\frac{1}{{{\beta ^2}}} = \frac{{16}}{{45}}\frac{{{e^4}\hbar }}{{m_e^4{c^7}}}$, where $m_e$ is the electron mass. In this context, we also recall the currently accepted upper limit for the photon mass, that is, ${m_\gamma } \sim 2 \times {10^{ - 16}}$ eV. Thus, for the QED case, from equation (\ref{E-H165}) it follows that the second term on the right-hand side it would be detectable in long-range distances ($\sim {10^9}$ m). In other words, we see that detectable corrections induced by vacuum polarization with a mass term would be present at low energy scales.

From equation (\ref{E-H165}), it clearly follows that the interaction energy between heavy charged charges, at leading order in $\beta$, is not finite at the origin. Motivated by this, one may consider the above calculation in a non-commutative geometry, based in findings of our previous studies \cite{Gaete-Helayel1,Gaete-Helayel2}. In such a case, the electric field at leading order in $\beta^2$ and $m^2$, takes the form
\begin{equation}
{E_i} = \left[ {{{\left( {1 + \frac{{{m^2}}}{\Delta }} \right)}^{ - 1}} + \frac{3}{{2{\beta ^2}}}{{\bf \Pi} ^2} - \frac{{6{m^2}}}{{{\beta ^2}}}{{\bf \Pi ^2}}\frac{1}{\Delta }} \right]{\partial _i}\left( { - \frac{{{e^{\theta {\nabla ^2}}}{\delta ^{\left( 3 \right)}}\left( {\bf x} \right)}}{{{\nabla ^2}}}} \right),\label{E-H170}
\end{equation}
where it may be recalled that we are now replacing the source ${\delta ^{\left( 3 \right)}}\left( {{\bf x} - {\bf y}} \right)$ by the smeared source  ${e^{\theta {\nabla ^2}}}{\delta ^{\left( 3 \right)}}\left( {{\bf x} - {\bf y}} \right)$, with $\theta$ the parameter non-commutative. Now, making use of equation (\ref{Vac85}), we readily find that
\begin{equation}
{{\cal A}_0}\left( {t,{\bf r}} \right) = {\cal A}_0^{\left( 1 \right)}\left( {t,{\bf r}} \right) + {\cal A}_0^{\left( 2 \right)}\left( {t,{\bf r}} \right) + {\cal A}_0^{\left( 3 \right)}\left( {t,{\bf r}} \right).  \label{E-H175}  
\end{equation}
The term ${\cal A}_0^{\left( 1 \right)}\left( {t,{\bf r}} \right)$ was first calculated in \cite{Gaete-Spallucci}, we can, therefore, write only the result:
\begin{equation}
{\cal A}_0^{\left( 1 \right)}\left( {t,{\bf r}} \right) = q\frac{{{e^{{m^2}\theta }}}}{{4\pi }}\frac{1}{r}\left[ {{e^{ - mr}} - \frac{1}{{\sqrt \pi  }}\int_{{\raise0.7ex\hbox{${{r^2}}$} \!\mathord{\left/
 {\vphantom {{{r^2}} {4\theta }}}\right.\kern-\nulldelimiterspace}
\!\lower0.7ex\hbox{${4\theta }$}}}^\infty  {du} \frac{1}{{\sqrt u }}{e^{ - u - \frac{{{m^2}{r^2}}}{{4u}}}}} \right] - q\frac{m}{{4\pi }}{e^{{m^2}\theta }}. \label{E-H180a}  
\end{equation}
While the terms ${\cal A}_0^{\left( 2 \right)}\left( {t,{\bf r}} \right)$ and ${\cal A}_0^{\left( 2 \right)}\left( {t,{\bf r}} \right)$, after some manipulation, can be brought to the form
\begin{equation}
{\cal A}_0^{\left( 2 \right)}\left( {t,{\bf r}} \right) = \frac{{12{q^3}}}{{{\beta ^2}{\pi ^{{\raise0.5ex\hbox{$\scriptstyle 3$}
\kern-0.1em/\kern-0.15em
\lower0.25ex\hbox{$\scriptstyle 2$}}}}}}{\hat n^i}\int_0^x {d{u^i}} \frac{1}{{{u^6}}}{\gamma ^3}\left( {{\textstyle{3 \over 2}},{\textstyle{{{u^2}} \over {4\theta }}}} \right), \label{E-H180b}
\end{equation}
\begin{equation}
{\cal A}_0^{\left( 3 \right)}\left( {t,{\bf r}} \right) = \frac{{3{m^2}{q^3}}}{{{\beta ^2}{\pi ^{{\raise0.5ex\hbox{$\scriptstyle 5$}
\kern-0.1em/\kern-0.15em
\lower0.25ex\hbox{$\scriptstyle 2$}}}}}}{\hat n^i}\int_0^x {d{u^i}} \frac{1}{{{u^4}}}{\gamma ^2}\left( {{\textstyle{3 \over 2}},{\textstyle{{{u^2}} \over {4\theta }}}} \right)\left[ {\frac{{4\theta }}{{{u^2}}}\gamma \left( {{\textstyle{3 \over 2}},{\textstyle{{{u^2}} \over {4\theta }}}} \right) - \gamma \left( {{\textstyle{1 \over 2}},{\textstyle{{{u^2}} \over {4\theta }}}} \right)} \right], \label{E-H180c}
\end{equation}
where $\gamma \left( {{\raise0.5ex\hbox{$\scriptstyle 3$}
\kern-0.1em/\kern-0.15em
\lower0.25ex\hbox{$\scriptstyle 2$}},{\raise0.5ex\hbox{$\scriptstyle {r^2 }$}
\kern-0.1em/\kern-0.15em
\lower0.25ex\hbox{$\scriptstyle {4\theta }$}}} \right)$ is the lower incomplete Gamma function defined by 
$\gamma \left( {{\raise0.5ex\hbox{$\scriptstyle a$}
\kern-0.1em/\kern-0.15em
\lower0.25ex\hbox{$\scriptstyle b$}},x} \right) \equiv \int_0^x {\frac{{du}}{u}} u^{{\raise0.5ex\hbox{$\scriptstyle a$}
\kern-0.1em/\kern-0.15em
\lower0.25ex\hbox{$\scriptstyle b$}}} e^{ - u}$. 

Inserting these expressions in equation (\ref{Vac80}), we finally obtain the static potential for two opposite 
charges $q$ located at $\bf 0$ and $\bf L$ as
\begin{eqnarray}
V &=&  - \frac{q}{{4\pi }}\frac{{{e^{{m^2}\theta }}}}{L}\left[ {{e^{ - mL}} - \frac{1}{{\sqrt \pi  }}\int_{{\raise0.7ex\hbox{${{L^2}}$} \!\mathord{\left/
 {\vphantom {{{L^2}} {4\theta }}}\right.\kern-\nulldelimiterspace}
\!\lower0.7ex\hbox{${4\theta }$}}}^\infty  {du} \frac{1}{{\sqrt u }}{e^{ - u - \frac{{{m^2}{u^2}}}{{4u}}}}} \right]
 - \frac{{12{q^4}}}{{{\beta ^2}{\pi ^{{\raise0.5ex\hbox{$\scriptstyle 3$}
\kern-0.1em/\kern-0.15em
\lower0.25ex\hbox{$\scriptstyle 2$}}}}}}{{\hat n}^i}\int_0^L {d{u^i}} \frac{1}{{{u^6}}}{\gamma ^3}\left( {{\textstyle{3 \over 2}},{\textstyle{{{u^2}} \over {4\theta }}}} \right) \nonumber\\
 &-& \frac{{3{m^2}{q^4}}}{{{\beta ^2}{\pi ^{{\raise0.5ex\hbox{$\scriptstyle 5$}
\kern-0.1em/\kern-0.15em
\lower0.25ex\hbox{$\scriptstyle 2$}}}}}}{{\hat n}^i}\int_0^x {d{u^i}} \frac{1}{{{u^4}}}{\gamma ^2}\left( {{\textstyle{3 \over 2}},{\textstyle{{{u^2}} \over {4\theta }}}} \right)\left[ {\frac{{4\theta }}{{{u^2}}}\gamma \left( {{\textstyle{3 \over 2}},{\textstyle{{{u^2}} \over {4\theta }}}} \right) - \gamma \left( {{\textstyle{1 \over 2}},{\textstyle{{{u^2}} \over {4\theta }}}} \right)} \right], \label{E-H185}
\end{eqnarray}
which is finite for $L \to 0$. It is a simple matter to verify that in the limit  $\theta \to 0$ we recover our above result.

\section{Logarithmic correction}

We now want to extend what we have done to Euler-Heisenberg-like electrodynamics at strong fields. As already mentioned, such theories show a power behavior that is typical for critical phenomena \cite{Kleinert}. In such a case the Lagrangian density reads:
\begin{equation}
{\cal L} =  - \frac{1}{4}{F_{\mu \nu }}{F^{\mu \nu }} - \frac{{{b_0}}}{8}{F_{\mu \nu }}{F^{\mu \nu }}\log \left( {\frac{{{F_{\mu \nu }}{F^{\mu \nu }}}}{{4{\lambda ^2}}}} \right), \label{corr05}
\end{equation}
where $b_0$ and $\lambda$ are constants. In fact, by choosing ${b_0} = \frac{{{e^2}}}{{6{\pi ^2}}}$ and $\lambda  = \frac{{m_e^2{c^3}}}{{e\hbar }}$, we recover the Euler-Heisenberg electrodynamics at strong fields \cite{Kleinert}.

In the same way as was done in the previous section one can introduce an auxiliary field, $\xi$, to handle the logarithm in (\ref{corr05}). This leads to 
\begin{equation}
{\cal L} =  - \frac{1}{4}{\alpha}_1{F_{\mu \nu }}{F^{\mu \nu }} - {\alpha}_2{\left( {{F_{\mu \nu }}{F^{\mu \nu }}} \right)^2}, \label{corr10}
\end{equation}
where  ${\alpha}_1 = 1 - \frac{{{b_0}}}{2}\left( {1 + \log \xi } \right)$ and ${\alpha}_2 = \frac{{{b_0}\xi }}{{32{\lambda ^2}}}$.\\

A similar procedure can be used to manipulate the quadratic term in (\ref{corr10}). Accordingly, by introducing a second auxiliary field, $\eta$, one easily finds
\begin{equation}
{\cal L} =  - \frac{1}{4}{F_{\mu \nu }}\left( {{\alpha}_1 + 4{\alpha}_2\eta } \right){F^{\mu \nu }} + \frac{{{\eta ^2}}}{4}{\alpha}_2. \label{corr15} 
\end{equation}
By setting, $\sigma  = {\alpha _1} + 4{\alpha _2}\eta$, we then have        
\begin{equation}
{\cal L} =  - \frac{1}{4}\sigma {F_{\mu \nu }}{F^{\mu \nu }} + \frac{1}{{64{\alpha}_2}}{\left( {\sigma  - {\alpha}_1} \right)^2}. \label{corr20}
\end{equation}

It is once again straightforward to apply the gauge-invariant formalism discussed in the foregoing section. The canonical momenta read $\Pi ^\mu   =  - \sigma F^{0\mu }$, and at once we recognize the two primary constraints $\Pi ^0  = 0$ and ${\cal P}_\sigma \equiv \frac{{\partial L}}{{\partial \dot \sigma}} = 0$. The canonical Hamiltonian corresponding to (\ref{corr20}) is 
\begin{equation}
{H_C} = \int {{d^3}x\left\{ {{\Pi _i}{\partial ^i}{A_0} + \frac{1}{{2\sigma }}{{\bf \Pi} ^2} + \frac{\sigma }{2}{{\bf B}^2} - \frac{1}{{64{\alpha}_2}}{{\left( {\sigma  - {\alpha}_1} \right)}^2}} \right\}}. \label{corr25}
\end{equation}

Requiring the primary constraint $\Pi^{0}$ to be preserved in time, one obtains the secondary constraint $\Gamma _1  = \partial _i \Pi ^i  = 0$. In the same way, for the constraint ${\cal P}_\sigma$, we get the auxiliary field $\sigma$ as 
\begin{equation}
\sigma  = \left( {1 - \frac{{{b_0}}}{2}\left( {1 + \ln \xi } \right) + \frac{{{b_0}{{\bf B}^2}}}{{2{\lambda ^2}}}\xi } \right)\left[ {1 + \frac{{3{b_0}{{\bf B}^2}}}{{2{\lambda ^2}}}\frac{\xi }{{{{\left( {1 - \frac{{{b_0}}}{2}\left( {1 + \ln \xi } \right) + \frac{{{b_0}{{\bf B}^2}}}{{2{\lambda ^2}}}\xi } \right)}^3}}}} \right]. \label{corr30}
\end{equation}
Hence we obtain
\begin{equation}
{H_C} = \int {{d^3}x} \left\{ {{\Pi _i}{\partial ^i}{A_0} + \frac{1}{2}{{\bf \Pi} ^2} + \frac{{{b_0}}}{4}\left( {1 + \log \xi } \right){{\bf \Pi} ^2} - \frac{{3{b_0}\xi }}{{2{\lambda ^2}}}{{\bf \Pi} ^4}} \right\}. \label{corr35}
\end{equation}

As before, requiring the primary constraint ${\cal P}_\xi$ to be preserved in time, one obtains the auxiliary field $\xi$. In this case $\xi  = \frac{\lambda }{{6{{\bf \Pi} ^2}}}$. Consequently, we get
\begin{equation}
{H_C} = \int {{d^3}x} \left\{ {{\Pi _i}{\partial ^i}{A_0} + \frac{1}{2}\left( {1 + {b_0}} \right){{\bf \Pi} ^2} - \frac{{6{b_0}}}{{{\lambda ^2}}}{{\bf \Pi} ^4}} \right\}. \label{corr40}
\end{equation}

Following the same steps that led to equation (\ref{E-H145}) we find that 
\begin{equation}
\left\langle H \right\rangle _\Phi ^{\left( 1 \right)}  = \left\langle \Phi  \right|\int {d^3 x} \left\{ {\frac{1}{2}{\bf \Pi} ^2  - \frac{3}{{8\beta ^2 }}{\bf \Pi} ^4 } \right\}\left| \Phi  \right\rangle. \label{corr45} 
\end{equation}
It should be noted that this expression is similar to equation (\ref{E-H145}) in the limit $m \to 0$, except by the chaged sign in frot of the ${\bf \Pi} ^4$ term. Hence we see that the potential for two opposite charges in $\bf 0$ and $\bf L$ is given by
\begin{equation}
V =  - \frac{{{q^2}}}{{4\pi }}\frac{{{1}}}{L} - \frac{{{q^4}}}{{6040 {\beta ^2}\pi^2}}{\frac{1}{{{L^5}}}}. \label{corr50}
\end{equation}

\section{Final Remarks}

Finally, within the gauge-invariant but path-dependent variables formalism, we have considered the confinement versus screening issue for both massive Euler-Heisenberg-like and Euler-Heisenberg Electrodynamics in the approximation of the strong-field limit. 
Once again, a correct identification of physical degrees of freedom has been fundamental for understanding the physics hidden in gauge theories. Interestingly enough, their non-commutative version displays an ultraviolet finite static potential. The above analysis reveals the key role played by the new quantum of length in our analysis. In a general perspective, the benefit of considering the present approach is to provide unifications among different models, as well as exploiting the equivalence in explicit calculations, as we have illustrated in the course of this work.

\begin{acknowledgments}
It is a pleasure to thank J. A. Helay\"{e}l-Neto for helpful comments on the manuscript.
This work was partially supported by Fondecyt (Chile) grant 1130426 and DGIP (UTFSM) internal project USM 111458. \end{acknowledgments}

\end{document}